\documentclass[a4paper,12pt]{article}
\usepackage{graphicx}
\usepackage{epsfig}
\usepackage{color}
\usepackage{subfigure}

\begin{document}

\thispagestyle{empty}
\vspace*{2cm}
\centerline{\Large\bf Simulation of Proton and Iron Induced Extensive Air Showers}
\centerline{\Large\bf for Different Hadronic Interaction Models
\footnote{
 Presented at $7^{th}$ PANE Conference, 2010}}
\vspace*{1.5cm}
\centerline{{\sc \footnote{chabinthakuria@gmail.com} C.~Thakuria$^{1}$},and {\sc K.~Boruah $^{2}$}}
\bigskip
\centerline{\sl $^1$Tihu College,,}
\centerline{\sl Tihu, Assam, 781371, INDIA}
\vskip0.6truecm
\centerline{\sl $^2$Department of Physics,}
\centerline{\sl Gauhati University, Guwahati-781014, INDIA}

\vspace*{1.5cm}
\centerline{\bf Abstract}
\vspace*{0.2cm}
\noindent
The reliable simulation of extensive air showers induced by different primary particles (e. g. proton, iron, gamma etc.) is of great importance in high energy cosmic ray research. The CORSIKA~\cite{abs} is a standard Monte-Carlo simulation package to simulate the four dimensional evolution of Extensive Air Shower (EAS) in the atmosphere initiated by gamma, hadrons and nuclei. CORSIKA has different high energy interaction models like DPMJET, QGSJET, NEXUS, SIBYLL, VENUS and EPOS which are based on different theoretical frameworks. The influence of different hadronic interaction models, viz., QGSJET and DPMJET on the lateral distribution functions (LDF) and muon to electron ratio of cosmic ray EAS induced by $\mbox{10}^{17}$ {\it eV} to $\mbox{10}^{20}$ {\it eV} proton and iron primaries are explored in this work.
\noindent

\vspace*{0.2cm}


\vfill

\newpage
\pagenumbering{arabic}

\section*{\normalsize\bf 1.~Introduction}
\label{intro}

The flux of cosmic ray particles extends to energies far beyond the reach of any Earth-based particle accelerator. Again direct observation of the first interaction of UHE cosmic rays in the upper atmosphere is impossible due to very low flux. The information on the first interactions can be derived from the obtained characteristics of primary induced cascades of secondary particles, known as extensive air showers(EAS); which requires ground based large detector setup. So the interpretation of EAS characteristics for primary energy above the energy attained by man made accelerator is dominated by model predictions and detailed simulations. Differences and  uncertainitties in these hadronic interaction models play the crucial role in the prediction of EAS characteristics. For air shower interpretation the understanding of multiparticle production in hadronic interactions with small momentum transfer is essential~\cite{one}. Due to energy dependence of the strong coupling constant as, soft interactions cannot be calculated within QCD using perturbation theory. Instead, phenomenological approaches have been introduced in different models incorporated in CORSIKA.
           Study of LDF(Lateral Distribution Function) for muons and electrons/positrons and photons is important for low energy hadronic model dependence study, especially LDF of muons.~\cite{three} . Again shower maximum prediction varies with defferent interaction models. Models that predict a higher scattering cross-section, higher multiplicity produces shower maximum at smaller slant depth.
           In the energy range  $\mbox{10}^{17}${\it eV} to $\mbox{10}^{20}${\it eV} lots of ambiguities regarding the composition of primary cosmic rays are prevailed. Therefore simulations for extreme assumptions for primary (viz. protons and iron nuclei) are done. Corresponding predictions for different observables are calculated for different interaction models. The measured data should be in between the predictions obtained for the extreme assumptions. If the data are outside the proton-iron range for an observable, there is is an indication for  an incompatibility of the particular hadronic interaction model with the observed values.~\cite{icrc/227/2009}.

\section*{\normalsize\bf Hadronic Interaction Model:}
\label{models}
In order to explain hadron-hadron/nucleus-nucleus collision above the attainable energy in accelerator one has to heavily rely on proposed hadronic interaction models.\textbf{ QGSJET (Quark Gluon String with JETs):}~\cite{five} is the most successful model in describing the high energy hadronic interactions. This model offered relatively easy approach to the simulation of cosmic ray interactions at higher energies, as well as ensured a good agreement to the accelerator data at lower energies~\cite{six}. This QGS based model treats the hadronic and nuclear collisions in the framework of Gribovs reggeon approach as multiple scattering processes, where individual scattering contributions are described phenomenologically as Pomeron exchanges. The Pomeron corresponds to microscopic partons (quark and gluon) cascades, which mediate the interaction between the projectile and the target hadrons, and consists of two parts: soft and semihard Pomerons. So in this model the appropriate cross sections for the inelastic interaction between hadrons i and j are calculated using the expression [1],
\vspace*{0.2cm}

  $ \chi_{ij}$$(s, b)$ $= \chi_{ij}^{s} (s, b) + \chi_{ij}^{h} (s, b)$  \hfill                               [1]
\vspace*{0.2cm}

where $\chi_{ij}^{s} (s, b)$ and $\chi_{ij}^{h} (s, b)$ represent the soft and semihard pomerons respectively. s is the centre of mass energy and b is the impact parameter.
\newpage
\noindent The total cross section in this model is given by the formula [2],
\vspace*{0.2cm}

$\sigma_{ij}^{t}(s)=\frac{1}{e_{ij}}\int d^{2}b\left[1-exp\left\{ −e_{ij}(\chi_{ij}^{s}(s, b) + \chi_{ij}^{h} (s, b))\right\} \right]$ \hfill [2]
\vspace*{0.2cm}

where $e_{ij}$ is so-called the shower enhancement co-efficient, for pp interactions epp = 1.5.
\vspace*{0.2cm}

\textbf { DPMJET (Dual Parton Model with JETs):} is based on the two components Dual Parton Model~\cite{seven} and contains multiple soft chains as well as multiple minijets. As QGSJET, it relies on the Gribov-Regge theory~\cite{eight, gribov} and the interaction is described by multi-Pomeron exchange. Soft processes are described by a supercritical Pomeron, while for hard processes additionally hard Pomerons are introduced. High mass diffractive events are described by triple Pomerons and Pomeron loop graphs, while low mass diffractive events are modelled outside the Gribov-Regge formalism. The total cross section in this model is given by [3],
\vspace*{0.2cm}

$\sigma_{ij}^{t}(b,s)=4\pi\intop_{0}^{\inf}bdb\left[1-exp(\chi_{ij}(s,b)\right]$ \hfill  [3]
\vspace*{0.2cm}

where $\chi_{ij}(s,b)$ represents total Pomeron in this model.

\section*{\normalsize\bf Method}
\label{meth}

Here we have used CORSIKA-6735 Simulation code to generate EAS using two hadronic interaction models viz. QGSJET-01 and DPMJET. We have considered two primary masses (proton and iron) and four primary energies($\mbox{10}^{17}${\it eV},$\mbox{10}^{18}${\it eV},$\mbox{10}^{19}${\it eV} and $\mbox{10}^{20}${\it eV}). There are altogether 16 sets of events with 1000 showers each. Simulations are performed for vertical showers. GHEISHA sub routine and THIN options are selected for flat horizontal detector. Data are first extracted by the inbuilt FORTRAN program and then these are analyzed using C++ program in the ROOT environment.

\begin{figure}[h]
\subfigure[Fig.1: Number of muons to number of electron Vs  primary energy]{
\includegraphics[width=0.5\textwidth]{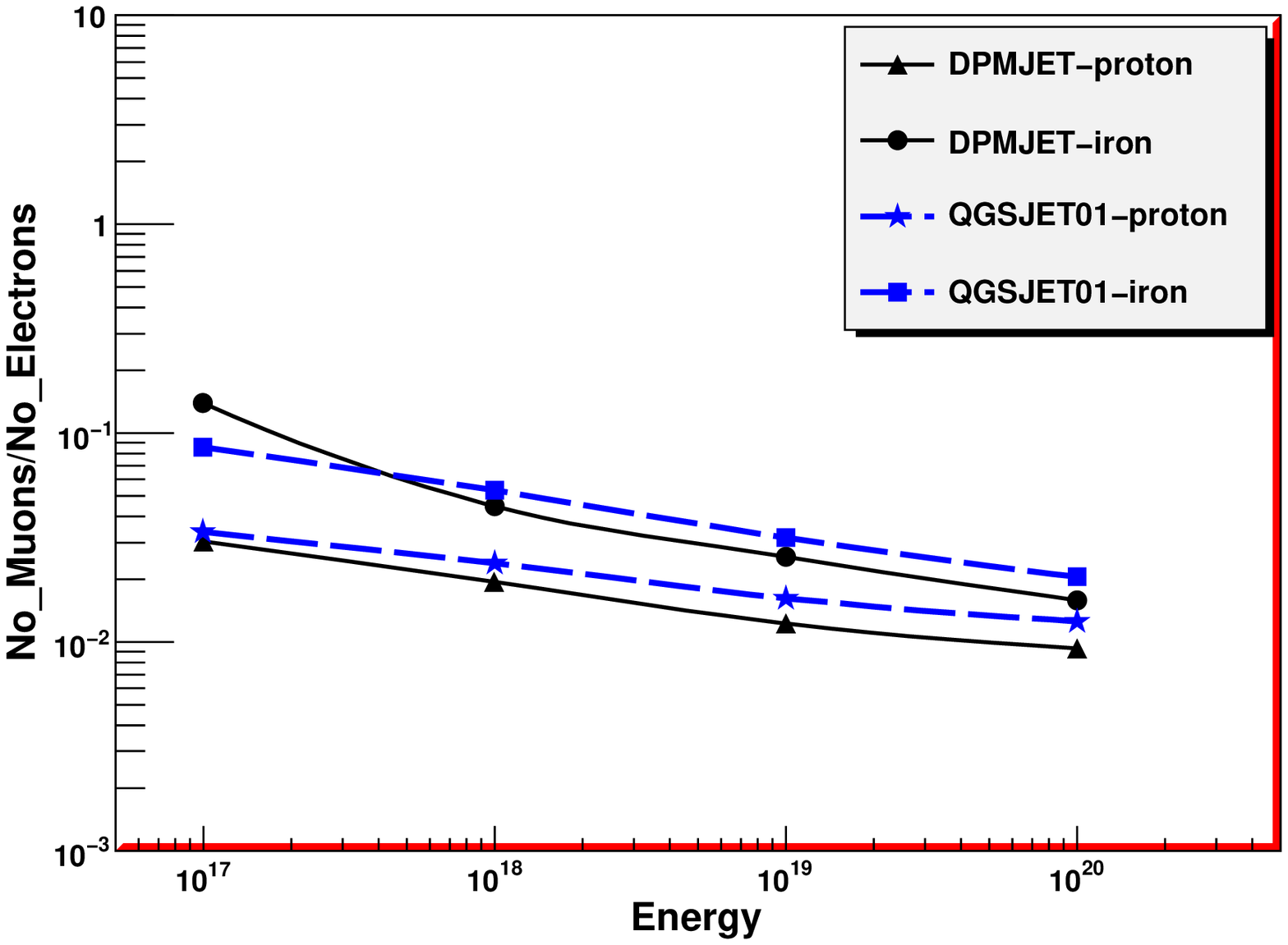}
 \label{Mu_El}          }
\subfigure[Fig.2:Average $X_{max}$ depth Vs log of primary energy]{
\includegraphics[width=0.45\textwidth]{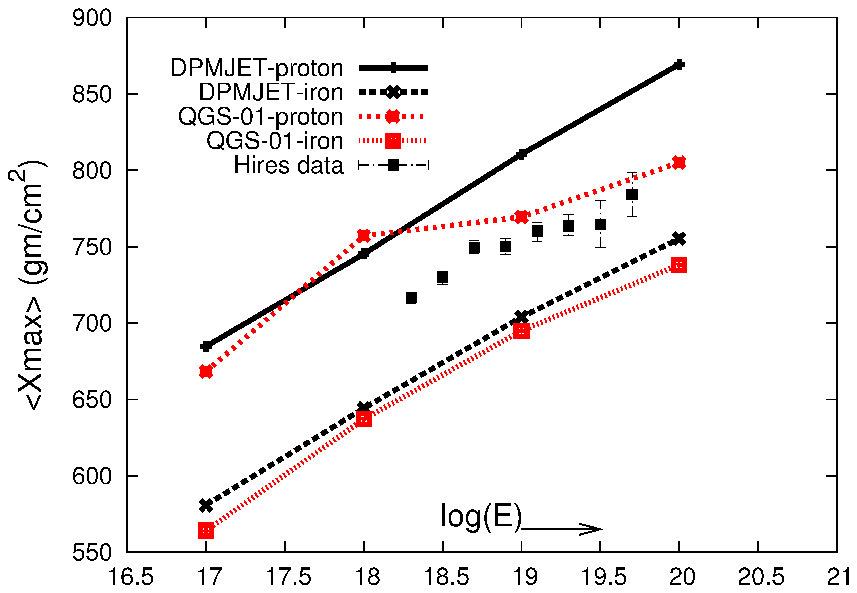}
 \label{Xmax}          }
\end{figure} 

\section*{\normalsize\bf Results:}
\label{res}

In order to study model dependence, we have chosen the following parameters:

                 ({\bf i}) the ratio of total number of muons to primary energy as a function of primary energy are plotted in Fig.1 The predictions by the DPMJET model for both the primaries have higher values except for proton induced shower at $\mbox{10}^{17}${\it eV}. Slopes of the curves for DPMJET and QGSJET are almost equal.

                ({\bf ii}) Average depth of shower maximum $<X_{max}>$ is plotted as a function of logarithm of primary energy as shown in Fig.2. The secondary data from Hires experimental setup are also superimposed into it. It is observed that irrespective of High Energy Hadronic Interaction Models considered, data are in between the predictions by proton induced showers and the same by iron induced showers.

               ({\bf iii}) Ratio of muon no to electron number as functions of primary energy and shower size are shown in Fig.3 \& Fig.4. Both these figures show 
that muon number is more for iron induced showers as compared to that produced for proton induced showers. However there is no significant difference so far as the 
two models are concerned.

                ({\bf iv}) Lastly, the density of muons and electrons at different core distance are calculated from the simulated data in 2m step and their ratio is plotted as function of core distance for proton and iron primaries. Fig.5 shows these results for two hadronic interaction models considered here. It is seen that there are more fluctuations at larger distances and no significant model dependence in all the cases. 
         
\begin{figure}[ht]
\subfigure[Fig.3:Number of muons divided by the primary energy Vs primary energy]{
\includegraphics[width=0.45\textwidth]{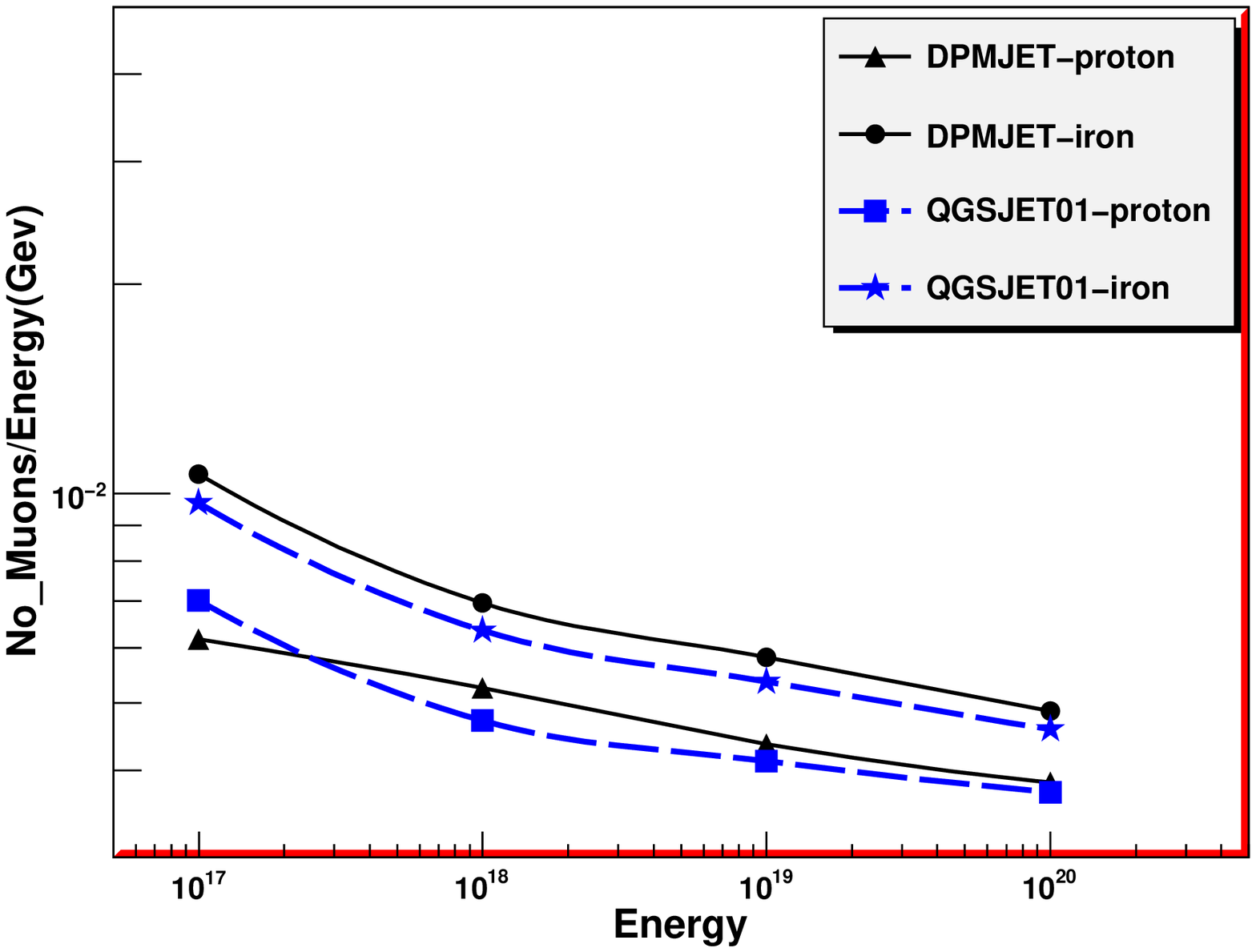}
\label{Muon_En}    }
\subfigure[Fig.4:Number of muons as function of Shower Size]{
 \includegraphics[width=0.45\textwidth]{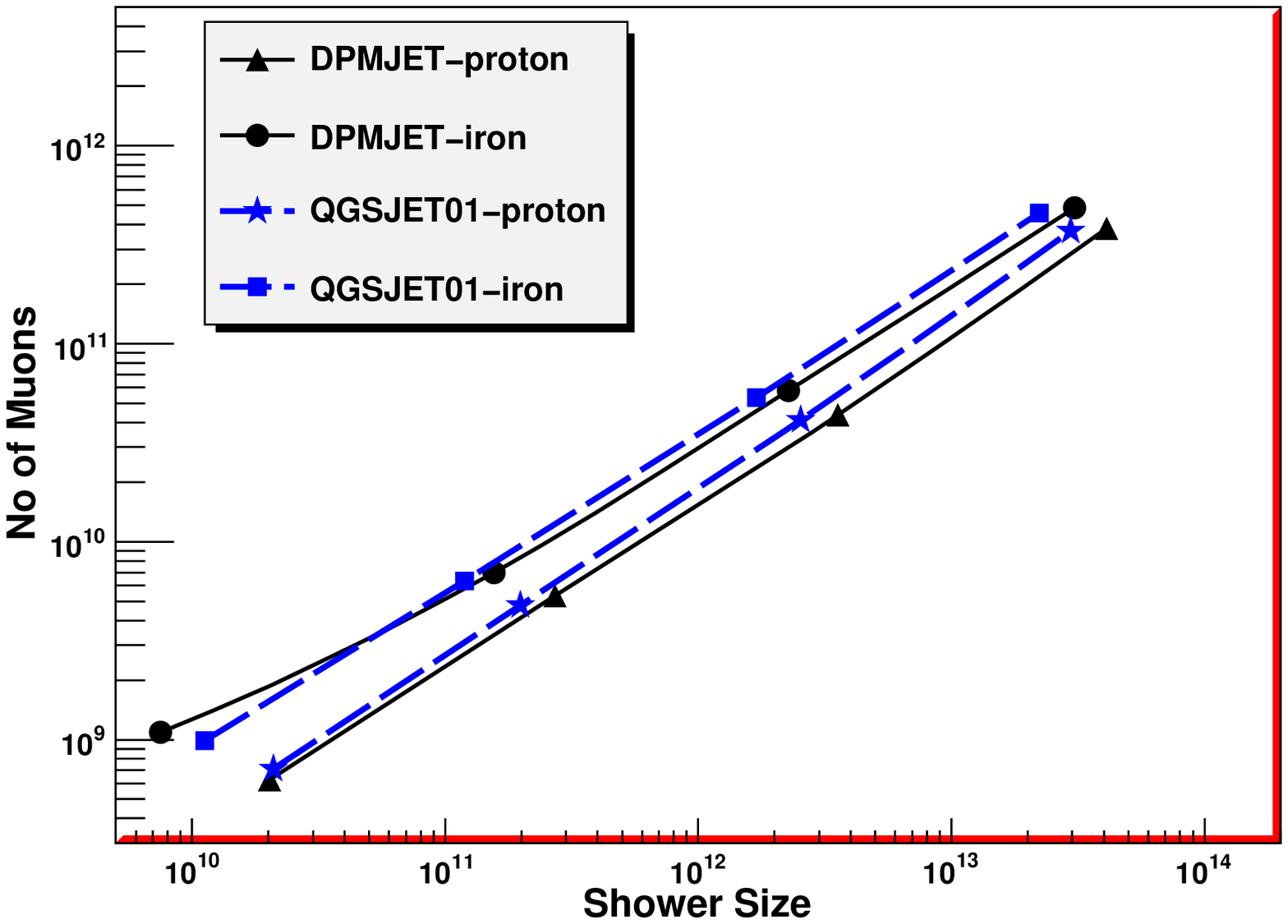}
\label{Mu_Sh}      }
\end{figure}

\begin{figure*}[ht]

 \includegraphics[width=1.0\textwidth]{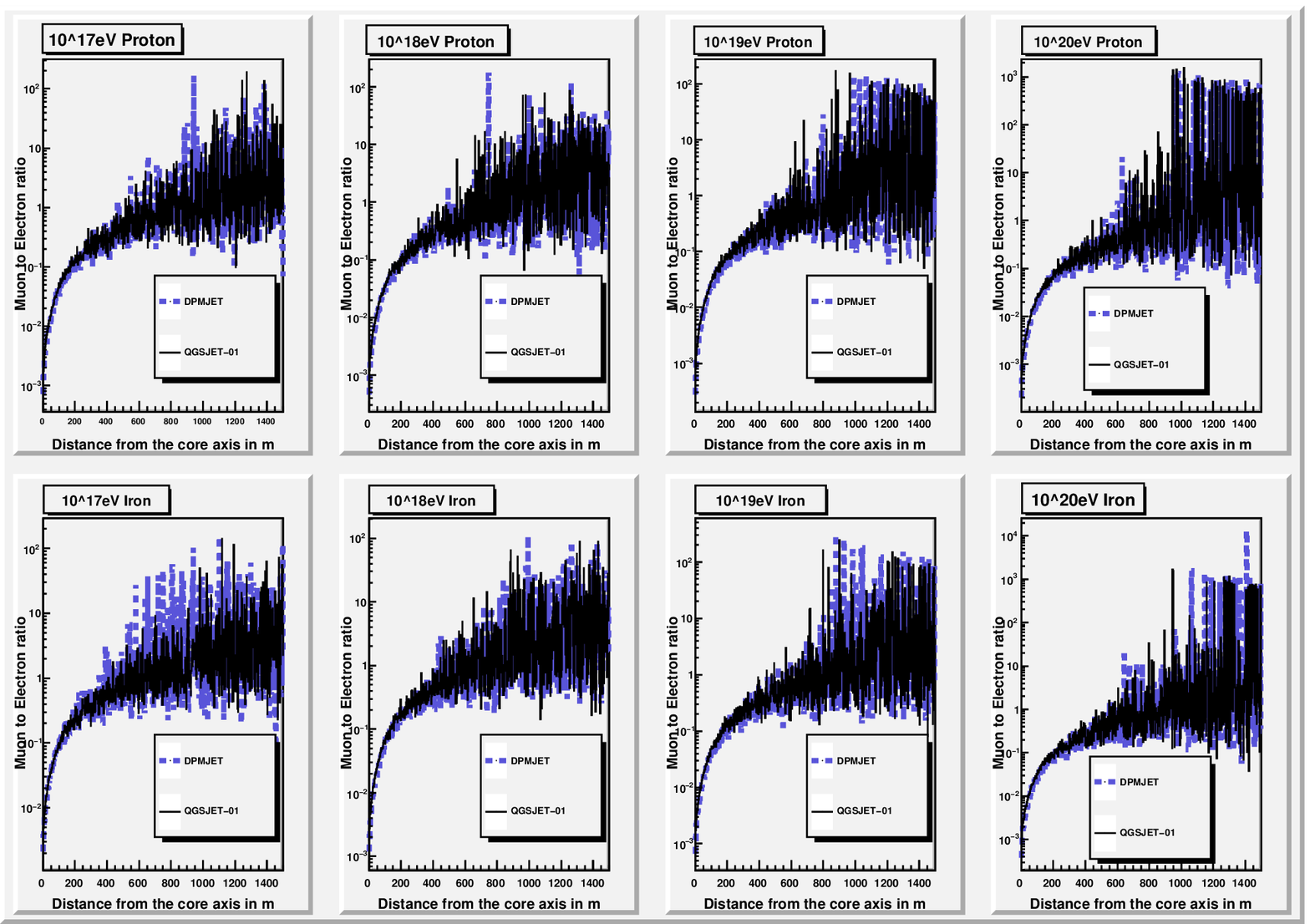}

 \figurename{ 5: Lateral distribution of muon to electron ratio for proton and iron primaries} 
\label{Latr}
\end{figure*}

\section*{\normalsize\bf Conclusion}
\label{concl}

 From the study of the different parameters it is found that the ratio of muon number to primary energy is best suited for distinguishing the effect of interaction model. DPMJET is found to produce more muons per {\it GeV} primary enegy compared to QGSJET-01 model.  
 
\vspace*{0.5cm}

\noindent {\bf Acknowledgements.}
The authors thankfully acknowledge U. D. Goswami of Dibrugarh University, India for his help and advice; and UGC for computational infrastructure support under SAP.


\vfill\eject

\end{document}